\documentclass[twocolumn]{aa}

\usepackage{graphicx}
\usepackage{txfonts}
\usepackage{color}

\begin{document}

\title{An analysis of the temperature structure of galaxy clusters by means of the thermal Sunyaev-Zel'dovich effect}

\author{Prokhorov, D. A.\inst{1,2}, Dubois, Y.\inst{3}, and Nagataki, S.\inst{2}}

\offprints{D.A. Prokhorov \email{prokhoro@iap.fr}}

\institute{Korea Astronomy and Space Science Institute, Hwaam-dong,
Yuseong-gu, Daejeon, 305-348, Republic of Korea
             \and
          Yukawa Institute for Theoretical Physics, Kyoto University,
          Kitashirakawa Oiwake-cho, Sakyo-ku, Kyoto, 606-8502, Japan
             \and
           Astrophysics, University of Oxford, Denys Wilkinson Building, Keble Road, Oxford, OX13RH, United Kingdom
            }

\date{Accepted . Received ; Draft printed: \today}

\authorrunning{D.A. Prokhorov et al.}

\titlerunning{An analysis of the temperature structure by means of the SZ effect}

\abstract
{}
{Measurements of the Sunyaev-Zel'dovich (hereafter SZ) effect
distortion of the cosmic microwave background provide us with an
independent method to derive the gas temperature of galaxy clusters.
In merging galaxy clusters the gas distribution is inhomogeneous
and, therefore, the method of temperature measuring based on the SZ
effect should be more relevant than that based on an X-ray emission
analysis. Here we study a method for measuring the gas temperature
in merging clusters by means of the SZ effect.}
{Our calculations of intensity maps of the SZ effect include
relativistic corrections considered within the framework of the
Wright formalism and utilize a cosmological numerical simulation of
a merging galaxy cluster evolved with its baryon physics. }
{We found that the gas temperature in merging clusters can be
measured by means of the ratio of the SZ intensity at a low
frequency (128 GHz) to that at a high frequency (369 GHz). This SZ
intensity ratio permits us to reveal prominent features of the
temperature structure caused by violent merger shock waves.
Therefore, measurements of the ratio of the SZ intensities are a
promising tool for measuring gas temperature in merging galaxy
clusters.}
{}

\keywords{galaxies: cluster: intracluster medium; relativistic
processes; cosmology: cosmic microwave background}

\maketitle

\section{Introduction}

Galaxy clusters are large structures in the Universe, with radii of
the order of a megaparsec. The space between galaxies in the
clusters is filled with low-density ($10^{-1}$--$10^{-3}$ cm$^{-3}$)
high temperature ($k_{\mathrm{B}} T\sim 1-10$ keV) electron-proton
plasma (for a review, see e.g., Sarazin 1986). Inverse Compton
scattering of hot free electrons in clusters of galaxies on the
cosmic microwave background (CMB) radiation field causes a change in
the intensity of the CMB radiation towards clusters of galaxies (the
Sunyaev-Zel'dovich effect, hereinafter the SZ effect; for a review,
see Sunyaev \& Zel'dovich 1980).

The SZ effect is important for cosmology and the study of clusters
of galaxies (for a review, see Birkinshaw 1999). It measures the
pressure of an electron population integrated along the line of
sight as long as free electrons are non-relativistic. Relativistic
effects are significant for high temperature plasmas in galaxy
clusters (see, e.g. Rephaeli 1995). A relativistically correct
formalism for the SZ effect based on the probability distribution of
the photon frequency shift after scattering was given by Wright
(1979) to describe the Comptonization process of soft photons by
mildly relativistic plasma.

Relativistic effects for the SZ effect permit us to measure the
temperature of intracluster plasma (see Pointecouteau et al. 1998;
Hansen et al. 2002). This method is more promising for measuring the
temperature of merging and/or distant clusters of galaxies than
those based on studies of an X-ray spectrum analysis (see
Pointecouteau et al. 1998). This is because the X-ray emission
traces the denser component (the X-ray emission is proportional to
$\int n_{\mathrm{e}}^2(l) dl$ (where $n_{\mathrm{e}}$ is the
electron number density and $dl$ is the integrated line of sight),
while the SZ effect is proportional to $\int n_{\mathrm{e}}(l) dl$)
and the cluster's X-ray surface brightness strongly decreases with
redshift while the SZ brightness is independent of redshift.

Using hydrodynamic simulations of galaxy clusters, Kay et al. (2008)
compare the temperatures derived from the X-ray spectroscopy and the
SZ effect. Their method for measuring the SZ temperature is based on
an equation (see Eq. 6 from Kay et al. 2008) which is only valid in
the Rayleigh-Jeans limit. As known relativistic effects on the CMB
intensity distortion are more significant at higher frequencies,
Colafrancesco \& Marchegiani (2010) conclude that to obtain detailed
information about the cluster temperature distribution one must use
high-frequency spectral observations of the SZ effect in the range
300-400 GHz and show that the SZ temperature can be extracted even
for cool non-merging galaxy clusters, such as the Perseus and Abell
2199 clusters.

In this paper, using the relativistically correct Wright formalism
and results of a cosmological numerical simulation of a cool distant
merging cluster, we show how the SZ temperature can be derived by
means of multi-frequency observations of the SZ effect. We calculate
intensity maps of the SZ effect at different frequencies and propose
to use the ratio of the SZ intensities at two frequencies to derive
the SZ temperature of galaxy clusters.

The layout of the paper is as follows. We describe the cosmological
numerical simulation of the cool distant merging cluster in Sect. 2.
We calculate the SZ intensity maps at different frequencies in the
framework of the Wright formalism in Sect. 3. We consider the ratio
of the SZ intensities at two frequencies to find a convenient method
for observing the SZ temperature of galaxy clusters in Sect. 4 and
present our discussions and conclusions in Sects. 5 and 6.

\section{Numerical simulations}

The galaxy cluster simulations of Dubois et al. (2010) which we use
in this paper are run with the Adaptive Mesh Refinement (AMR) code
RAMSES (Teyssier 2002). The evolution of the gas is followed using a
second-order unsplit Godunov scheme for the Euler equations. The
Riemann solver used to compute the flux at a cell interface is the
acoustic solver using a first-order MinMod total variation
diminishing scheme to reconstruct the interpolated variables from
their cell-centered values. Collisionless particles (dark matter,
stars and black hole particles) are evolved using a particle-mesh solver
with cloud-in-cell interpolation.

The simulations are performed using a re-simulation (zoom)
technique: the coarse region is a $128^3$ grid with
M$_{\mathrm{DM}}$ = $2.9\times10^{10}$ M$_{\odot}$ DM resolution in
a 80 h$^{-1}_{70}$ Mpc simulation box, where $h_{70}$ is the Hubble
constant in units of 70 km s$^{-1}$ Mpc$^{-1}$. This region contains
a smaller $256^3$ equivalent grid in a sphere of radius 20
h$^{-1}_{70}$ Mpc with M$_{\mathrm{DM}}$ = $3.6\times10^9$
M$_{\odot}$ DM resolution, which in turn encloses the final high
resolution sphere with radius 6 h$^{-1}_{70}$ Mpc, $512^3$
equivalent grid and M$_{\mathrm{DM}}$ = $4.5\times10^8$ M$_{\odot}$
DM resolution. The maximum level of refinement reached in this
simulation allows to resolve a minimum spatial scale of $1.19 \, \rm
h^{-1}_{70} kpc$.

A flat $\Lambda$CDM cosmology was assumed with total matter density
$\Omega_{m}=0.3$, baryon density $\Omega_b=0.045$, dark energy
density $\Omega_{\Lambda}=0.7$, fluctuation amplitude at $8 \,
 \rm h^{-1}_{70}$ Mpc $\sigma_8=0.90$ and Hubble constant H$_0$=70 km s$^{-1}$ Mpc$^{-1}$
that corresponds to the Wilkinson Microwave Anisotropies Probe 1
year best-fitting cosmology (Spergel et al. 2003).

\begin{figure}[ht]
\centering
\includegraphics[angle=0, width=7.5cm]{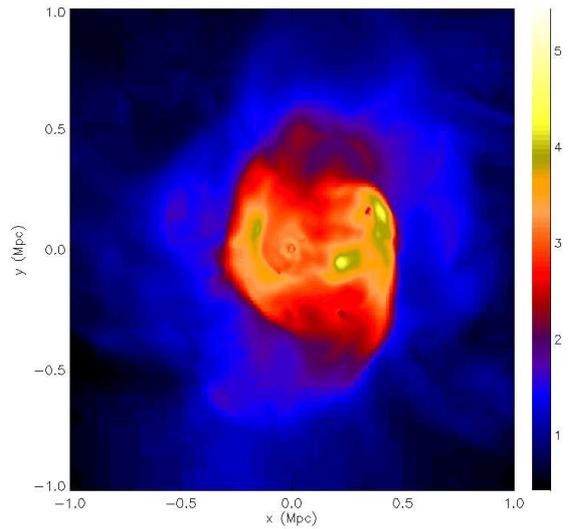}
\caption{The mass-weighted temperature map (in keV) of the simulated
cluster along the x-direction at z=0.74} \label{tem}
\end{figure}

The resimulated region tracks the formation of a galaxy cluster with
a 1:1 major merger occurring at z = 0.8. This z = 0.8 major galaxy
merger drives the cluster gas to temperatures twice the virial
temperature thanks to violent shock waves. The projected
mass-weighted temperature map is shown in Fig. \ref{tem}, where
x-direction is orthogonal to the merger plane.

Standard recipes for atomic cooling, heating from UV background and star formation in high density regions are included.
This simulation also includes a novel recipe for AGN feedback based on a kinetic-jet energy input that prevents the cooling catastrophe to occurs, the formation of unrealistic massive cooling flows, and regulates the star formation in the central galaxy (see Dubois et al. 2010 for further details on the simulation).

\section{The SZ effect from the simulated cluster}

In this section, the SZ intensity maps are calculated in the
framework of the relativistic Wright formalism for the simulated
galaxy cluster.

The CMB intensity change produced by the SZ effect by
non-relativistic electrons considered within the framework of the
Kompaneets approximation is (see for a review, Birkinshaw 1999):

\begin{equation}
\Delta I_{nr}(x) = I_{\mathrm{0}} g(x) y_{\mathrm{gas}} \label{Inr}
\end{equation}
where $x=h\nu/k_{\mathrm{b}} T_{\mathrm{cmb}}$, $I_{\mathrm{0}}=2
(k_{\mathrm{b}} T_{\mathrm{cmb}})^3 / (hc)^2$, and the spectral
function $g(x)$ is given by
\begin{equation}
g(x)=\frac{x^4 \exp(x)}{(\exp(x)-1)^2} \left(x
\frac{\exp(x)+1}{\exp(x)-1}-4\right).
\end{equation}
The subscript  $`nr' $ denotes the fact that Eq. (\ref{Inr}) was
obtained in the non-relativistic limit. The Comptonization parameter
$y_{\mathrm{gas}}$ is given by
\begin{equation}
y_{\mathrm{gas}}=\frac{\sigma_{\mathrm{T}}}{m_{\mathrm{e}}c^2} \int
n_{\mathrm{gas}} k T_{\mathrm{e}} dl
\end{equation}
where the line-of-sight integral extends from the last scattering
surface of the CMB radiation to the observer at redshift z=0,
$T_{\mathrm{e}}$ is the electron temperature, $n_{\mathrm{gas}}$ is
the number density of the gas, $\sigma_{\mathrm{T}}$ is the Thomson
cross-section, $m_{\mathrm{e}}$ the electron mass, $c$ the speed of
light, $k_{\mathrm{b}}$ the Boltzmann constant and $h$ the Planck
constant.

The CMB intensity change in the Wright formalism can be written in
the form proposed by Prokhorov et al. (2010), and is
\begin{equation}
\Delta I(x) = I_{\mathrm{0}}
\frac{\sigma_{\mathrm{T}}}{m_{\mathrm{e}}c^2} \int n_{\mathrm{gas}}
k_{\mathrm{b}} T_{\mathrm{e}} G(x, T_{\mathrm{e}}) dl, \label{form}
\end{equation}
where the spectral function $g(x)$ is changed to the generalized
spectral function $G(x, T_{\mathrm{e}})$ which depends explicitly on
the electron temperature.

The relativistic spectral function $G(x, T_{\mathrm{e}})$ derived in
the framework of the Wright formalism is given by
\begin{equation}
G(x, T_{\mathrm{e}})=\int^{\infty}_{-\infty} \frac{P_{1}(s,
T_{\mathrm{e}})}{\Theta(T_{\mathrm{e}})} \left(\frac{x^3 \exp(-3
s)}{\exp(x \exp(-s))-1}-\frac{x^3}{\exp(x)-1}\right) ds \label{G}
\end{equation}
where $\Theta(T_{\mathrm{e}}) = k_{\mathrm{b}} T_{\mathrm{e}}/
m_{\mathrm{e}}c^2$, and $P_{1}(s, T_{\mathrm{e}})$ is the
distribution of frequency shifts for single scattering (Wright 1979;
Birkinshaw 1999).

There are three basic spectral features that characterize the
thermal, non-relativistic SZ effect signal: a minimum of its
intensity is located at a dimensionless frequency $x_1=2.26$
($\nu$=128 GHz), the crossover frequency is $x_0$ = 3.83 ($\nu$=217
GHz), and a maximum of its intensity is located at a dimensionless
frequency $x_2=6.51$ ($\nu$ = 369 GHz).

Observations of the SZ effect close to the crossover frequency are
biased due to the presence of the kinematical SZ effect (for a
review, see Birkinshaw 1999), which is associated with the peculiar
velocity of the galaxy cluster. The unknown value of the peculiar
velocity limits the ability to measure the cluster temperature
directly through the displacement of the crossover frequency of the
SZ effect in the correct relativistic treatment (Colafrancesco et
al. 2009).

Prokhorov et al. (2010) show that the choice of frequencies $x_1 =
2.26$ and $x_2 = 6.51$ corresponding to minimum and maximum values
of the SZ intensity in the Kompaneets approximation is suitable to
analyze mildly relativistic electron populations. Since the
frequency of its maximum is located at 369 GHz, observations at this
high frequency should be promising to derive the temperature by
means of the SZ effect (Colafrancesco et al. 2010).

\begin{figure}[ht]
\centering
\includegraphics[angle=0, width=7.5cm]{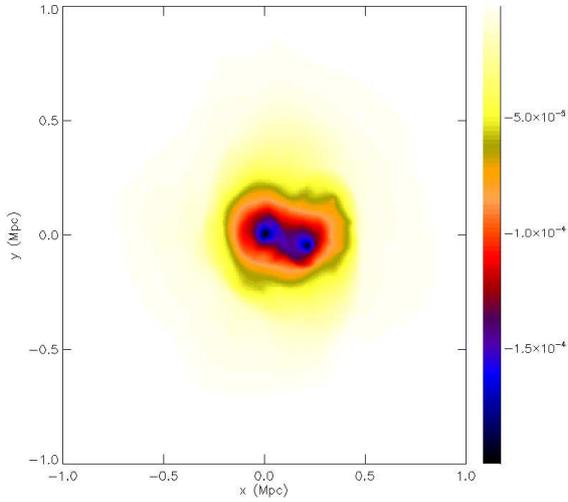}
\caption{The intensity map $I/I_0$ of the SZ effect at a frequency
128 GHz derived from the numerical simulation in the framework of
the Wright formalism.} \label{SZ128}
\end{figure}

\begin{figure}[ht]
\centering
\includegraphics[angle=0, width=7.5cm]{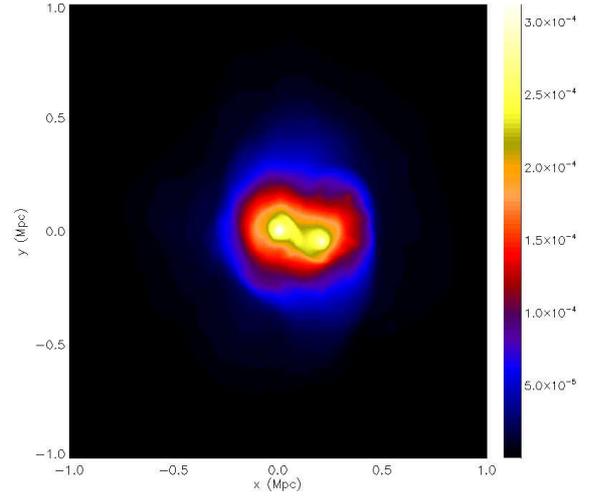}
\caption{The intensity map $I/I_0$ of the SZ effect at a frequency
369 GHz derived from the numerical simulation in the framework of
the Wright formalism.} \label{SZ369}
\end{figure}

To produce the SZ intensity maps at frequencies $x_1 = 2.26$ and
$x_2 = 6.51$ we use the 3D density and temperature maps for the
simulated galaxy cluster considered in Sect. 2. We calculated the SZ
effect using the Wright formalism. The intensity maps of the SZ
effect at these frequencies derived from the simulation maps of the
gas density and temperature are plotted in Figs. \ref{SZ128} and
\ref{SZ369}, respectively.

The morphologies of the SZ intensity simulated maps at frequencies
$x_1 = 2.26$ and $x_2 = 6.51$ are similar. This is because the SZ
effect from cool galaxy clusters can be approximately described in
the framework of the Kompaneets approximation. Note that the
spectral function $g(x)$ does not depend on gas temperature.
However, we now show that the gas temperature structure of a cool
galaxy cluster can be derived by means of a more accurate analysis.

\section{The ratio of the SZ intensity at frequency 128 GHz to that at frequency 369 GHz}

In this section, we show that the ratio of the SZ intensities at
frequencies 128 GHz to 369 GHz provides us with a convenient method
for measuring the SZ temperature of galaxy clusters.

Multi-frequency analysis of the SZ effect permits us to derive
temperature maps for galaxy clusters. Colafrancesco \& Marchegiani
(2010) fitted the six frequency SZ effect data ($\nu$ = 300, 320,
340, 360, 380 and 400 GHz) with a relativistic model of the thermal
SZ effect and assuming, for each experimental data point, an
uncertainty of 0.1\%. They present results of the fitting procedure
used to extract the cluster temperature from a set of simulated
spatially resolved spectroscopic SZ effect observations in different
bands of the spectrum. The Perseus and Abell 2199 clusters studied
in Colafrancesco \& Marchegiani (2010) are relaxed galaxy clusters,
here we present our results of an analysis of the temperature
structure of the simulated merging galaxy cluster.

Using the relativistic Wright formalism for the SZ effect we found
that the ratio of the relativistic spectral function $G(x_{1},
T_{\mathrm{e}})$ at frequency 128 GHz to that of function $G(x_2,
T_{\mathrm{e}})$ at frequency 369 GHz is a monotonic function of
temperature. For an isothermal galaxy cluster the ratio of the SZ
intensities is given by $\Delta I(x_{1})/\Delta I(x_{2})=G(x_{1},
T_{\mathrm{e}})/G(x_{2}, T_{\mathrm{e}})$. Note that the ratio of
the SZ intensities does not depend on temperature in the framework
of the Kompaneets formalism.

We have checked the obtained monotonical dependance using the
generalized Kompaneets equation derived by Challinor \& Lasenby
(1998) including relativistic effects. Using their Eq. (28) which is
valid for $k_{\mathrm{B}} T_{\mathrm{e}}<10$ keV we find that the
ratio of the SZ intensity at frequency 128 GHz to that at frequency
369 GHz equals

\begin{equation}
\frac{\Delta I(x_{1})}{\Delta
I(x_{2})}\approx-0.607-2.143\times\frac{k_{\mathrm{B}}
T_{\mathrm{e}}}{m_{\mathrm{e}} c^2}.
\label{ch}
\end{equation}

Comparing Eq. (\ref{form}) from this paper and Eq. (6) from Kay et
al. (2008) we find that the temperature derived from the ratio of
the SZ intensities at frequency 128 GHz to that at frequency 369 GHz
within the framework of the Wright formalism is equivalent the
Compton-averaged electron temperature for cool galaxy clusters for
which Eq. (28) from Challinor \& Lasenby (1998) is valid.

\begin{figure}[ht]
\centering
\includegraphics[angle=0, width=7.5cm]{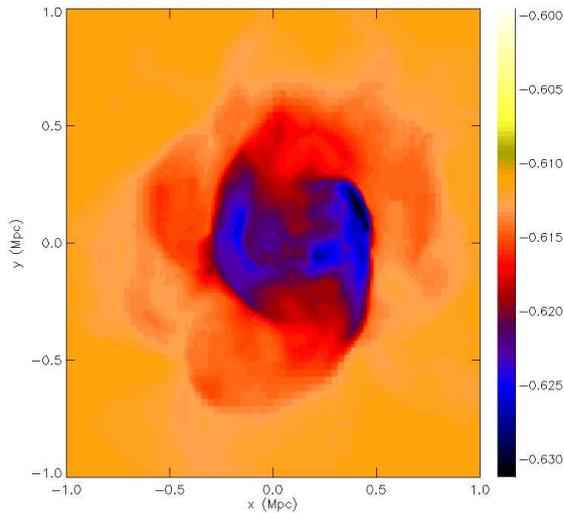}
\caption{The ratio of the SZ intensity at frequency 128 GHz to that
at frequency 369 GHz for the simulated cluster}
\label{ratio}
\end{figure}

The ratio of the SZ intensity maps at frequencies 128 GHz and 369
GHz derived within the framework of the Wright formalism (see Figs.
\ref{SZ128} and \ref{SZ369}) is shown in Fig. \ref{ratio}. We note
that regions of high temperature on the mass-weighted temperature
map (see Fig. \ref{tem}) corresponds to regions with low values of
the SZ intensity ratio as expected from Eq. (\ref{ch}).

\begin{figure}[ht]
\centering
\includegraphics[angle=0, width=7.5cm]{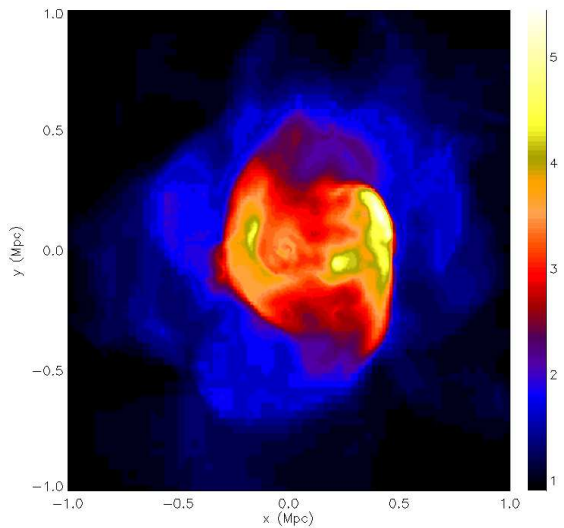}
\caption{Temperature (in keV) of the simulated cluster along the x
direction at z=0.74 derived from the ratio of the SZ intensities. }
\label{sztem}
\end{figure}

Using the monotonical dependance of the SZ intensity ratio on
temperature derived within the framework of the Wright formalism we
derive the SZ temperature map from the map of the ratio of the SZ
intensities at frequencies 128 GHz to 369 GHz. The SZ temperature
map is shown in Fig. \ref{sztem}.

The temperature maps in Figs. \ref{tem} and \ref{sztem} show high
temperature ``arc-like'' structures. The origin of these
``arc-like'' structures on the temperature maps is the major merger
of two galaxy clusters occurring at z=0.8. The evolution of the
mass-weighted temperature along the x-direction during its merging
phase is shown in Fig. 5 from Dubois et al. (2010) at different
redshifts z=0.88, z=0.74, and z=0.58. As shown by Dubois et al.
(2010) this major merger drives the cluster gas to temperatures
twice the virial temperature thanks to violent merger shock waves.

Figure \ref{sztem} shows that the SZ temperature is higher than the
mass-weighted temperature (see Fig. \ref{tem}). This is consistent
with the results of Kay et al. (2008).

We find, from Eq. (6), that a sensitivity of order of $\sim$ 0.3\%
of the predicted SZ effect is necessary to measure the SZ
temperature in cool galaxy clusters with a precision of 1 keV by
means of the SZ intensity ratio.

Our analysis shows that the ratio of the SZ intensities is a
promising approach for measuring the temperature structure in
merging galaxy clusters.

\section{Discussions}

A source of bias in the observations of the SZ effect could be
provided by a possibly relevant kinematic SZ effect. The kinematic
SZ effect arises from the bulk motion of the medium relative to the
CMB rest frame. In this section we propose a method to extract the
kinematical SZ effect from SZ observations and also discuss various
constraints on the ability of measuring the gas temperature in
distant merging galaxy clusters by means of X-ray observations.

\subsection{The kinematic SZ effect}

To study the kinematical SZ effect from SZ observations, Rephaeli \&
Lafav (1991) proposed to measure the SZ effect at the crossover
frequency. The contribution of the kinematical SZ effect is maximal
at the crossover while the thermal SZ effect in the Kompaneets
approximation equals zero. Using Eq. (28) from Challinor \& Lasenby
(1998), we calculate the ratio of the intensities of the
relativistic SZ effect to the kinematic SZ effect at the crossover
frequency. This ratio is given by

\begin{equation}
\frac{I_{\mathrm{rel}}(x=3.83)}{I_{\mathrm{kin}}(x=3.83)}\approx-0.23
\left(\frac{k_{\mathrm{b}} T_{\mathrm{e}}}{4 \mathrm{keV}}\right)^2
\frac{300 \mathrm{km/s}}{\mathrm{v}} \label{kin}
\end{equation}
where $\mathrm{v}$ is the peculiar velocity of a galaxy cluster and
300 km/s is the rms value of the peculiar velocity distribution of
galaxy clusters (see Giovanelli et al. 1998).

Equation \ref{kin} shows that the relativistic corrections to the
thermal SZ effect in galaxy clusters at the crossover frequency can
be significant and biases measurements of peculiar velocity by the
Rephaeli \& Lafav method, particulary in hot clusters with
temperature $\sim 10$ keV. Therefore, below we propose another
method to measure peculiar velocities. We note that the approximate
function described the SZ relativistic corrections taken from
Challinor \& Lasenby (1998) for $k_{\mathrm{B}} T_{\mathrm{e}}<10$
keV has two frequencies at which relativistic corrections equal
zero. The values of these frequencies are $x_{\mathrm{a}}=3.33$ and
$x_{\mathrm{b}}=8.02$. Measurements of SZ intensities at these
frequencies provides us with a method to derive the peculiar
velocity. Using Eq. (28) from Challinor \& Lasenby (1998), we find

\begin{equation}
\left(\int n_{\mathrm{e}}
dl\right)\frac{\mathrm{v}}{c}=\frac{m_{\mathrm{e}} c^2}{I_{0}
\sigma_{\mathrm{T}}}\times\frac{\Delta I(x_{\mathrm{a}})
g(x_{\mathrm{b}})-\Delta I(x_{\mathrm{b}})
g(x_{\mathrm{a}})}{h(x_{\mathrm{a}}) g(x_{\mathrm{b}}) -
h(x_{\mathrm{b}}) g(x_{\mathrm{a}})} \label{SZkin}
\end{equation}
where the spectral function h(x) describing the kinematic SZ effect
is

\begin{equation}
h(x)=\frac{x^4 \exp(x)}{(\exp(x)-1)^2}.
\end{equation}

The optical depth of the gas to Compton scattering can be obtained
from spectral and spatial X-ray observations, and can be determined
accurately in relaxed galaxy clusters. The proposed method to derive
the peculiar velocity of a galaxy cluster based on Eq. (\ref{SZkin})
is not affected by biases due to relativistic corrections of the
thermal SZ effect.

It is important that measurements of the value of $\left(\int
n_{\mathrm{e}} dl\right) \mathrm{v}/c$ permit us to extract
contributions of the kinematical SZ effect from SZ observations at
frequencies of 128 GHz and 369 GHz, since the contribution of the
kinematical SZ effect is proportional to $\left(\int n_{\mathrm{e}}
dl\right)\mathrm{v}/c$.

\subsection{Constraints on the ability of measuring the gas
temperature via X-ray observations}

\begin{figure}[ht]
\centering
\includegraphics[angle=0, width=7.5cm]{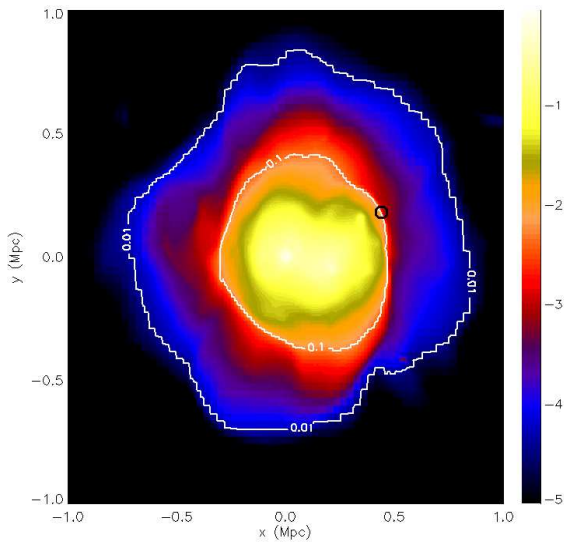}
\caption{The normalized X-ray surface brightness map of the
simulated cluster in the [2.0–-10.0 keV] band in logarithmic scale.
The region corresponding to the highest temperature in Fig.
\ref{sztem} is shown by a black circle. } \label{surf}
\end{figure}

To compare methods for measuring the gas temperature in merging
clusters based on the SZ effect with that based on X-ray emission,
we produce the X-ray surface brightness map of thermal
bremsstrahlung emission for the simulated cluster. For the sake of
illustration, the normalized X-ray surface brightness map of the
simulated cluster in the [2.0–-10.0 keV] band in logarithmic scale
is shown in Fig. \ref{surf}, which can be obtained by modern X-ray
satellites, such as XMM-Newton, Chandra, and Suzaku. The region
corresponding to the highest temperature in Fig. \ref{sztem} is
shown by a black circle in Fig. \ref{surf}. The X-ray surface
brightness of this region is two orders of magnitude smaller than
the maximal surface brightness value in Fig. \ref{surf}, while the
SZ intensity (see contours of the SZ intensity superimposed on the
X-ray surface brightness map, the inner contour represent 10\% of
the maximum of the SZ intensity and the outer contour represents 1\%
of the maximum of the SZ intensity) of this region is one order of
magnitude smaller than the maximal SZ intensity value in Figs.
\ref{SZ128} and \ref{SZ369}. Therefore the X-ray emission traces the
denser component while the SZ effect provides us with a method to
study more rarified gas regions.

The Abell 3376 cluster at redshift z=0.046 is a cool nearby cluster
and provides us with an example of galaxy clusters where a shock has
not been revealed by X-ray observations with the XMM-Newton X-ray
satellite, but instead it has been detected by radio observations
with the Very Large Array telescope (Bagchi et al. 2006). This is
because the X-ray surface brightness in the region of the east
ringlike structure is very weak and does not permit us to detect
this shock.

Since the X-ray surface brightness decreases with redshift as
$1/(1+z)^4$, the X-ray surface brightness for the simulated cluster
at z=0.74 decreases by a factor $\sim$9 compared with the same
cluster at z=0. The decrease of the X-ray surface brightness with
redshift constrains the ability of measuring the gas temperature in
distant galaxy clusters by means of X-ray observations.

\section{Conclusions}

Merging galaxy clusters are an interesting astrophysical laboratory
for studying gasdynamic processes. Mergers of galaxy clusters are
very energetic astrophysical events in which huge gravitational
energy is released. In the course of a merger, a significant portion
of this energy, which is carried by the gas, is dissipated by merger
shock waves. This leads to a heating of the gas to higher
temperatures. Studying the temperature structure of merging galaxy
clusters is important to reveal heated gas regions associated with
merger shock waves.

Studying X-ray spectra provides us two independent approaches to
determine the gas temperature in galaxy clusters, this is because
the bremsstrahlung continuum spectrum depends on the gas temperature
and ratios of emission line fluxes are functions of the gas
temperature (e.g. the ratio of the He-like to H-like iron line
fluxes, see e.g. Prokhorov et al. 2009). An analysis of the SZ
effect permits us to determine the gas temperature by studying
deviations of the SZ intensity spectrum from that derived within the
framework of the Kompaneets approximation. Colafrancesco \&
Marchegiani (2010) show that the method based on the SZ intensity
deviations can be interesting for measuring the gas temperature even
in cool clusters, such as the Perseus and Abell 2199 clusters, if
uncertainties of observational data are 0.1\% - 1\%. In this paper,
using the relativistically correct Wright formalism and results of
3-D hydrodynamic numerical simulations of a cool distant merging
cluster, we show how the SZ temperature can be derived by means of
multi-frequency observations of the SZ effect.

To produce a realistic cool distant merging cluster we use a zoom
cosmological simulation from Dubois et al. (2010). We find that the
simulated cluster undergoes a violent merger at z=0.8, which drives
the cluster gas to higher temperatures. The temperature map of the
simulated cluster becomes very inhomogeneous because of the merger
activity (see Fig. \ref{tem}). We consider a method for measuring
the temperature structure in this merging simulated cluster by means
of distortions of the SZ intensity map due to relativistic effects.

We incorporate the relativistic Wright formalism for modeling the SZ
effect in the numerical simulation using the algorithm proposed by
Prokhorov et al. (2010). We calculate the SZ intensity maps at low
(128 GHz) and high (369 GHz) frequencies and find these SZ intensity
maps look similar as expected for a cool galaxy cluster. To provide
a method for measuring the temperature structure in cool merging
galaxy clusters we propose to use the ratio of the SZ intensities at
two frequencies.

We calculate the ratio of the SZ intensity at a low frequency (128
GHz) to that at a high frequency (369 GHz) for the simulated merging
cluster and show that this ratio is a promising method to measure
the SZ temperature in galaxy clusters. The calculated map of the SZ
temperature shows that main features (such as ``arc-like''
structures) of the temperature structures of merging clusters may be
revealed by using this method.

The next generation SZ effect experiments discussed in Colafrancesco
\& Marchegiani (2010) are needed to reach the required high
sensitivity with the purpose of being independent in measuring gas
temperature in galaxy clusters. Experimental configurations which
are basically the same as those of the Millimetron project (see
http://www.asc.rssi.ru/millimetron/) will have the power to measure
the gas temperature via the SZ effect in any cluster, including cool
ones.

\begin{acknowledgement}
We are grateful to Florence Durret for valuable discussions of X-ray
emission of the Abell 3376 cluster and to Sergio Colafrancesco for
valuable discussions of experimental configurations for Millimetron.
\end{acknowledgement}


\begin{thebibliography}{99}
\bibitem{Bagchi 2006}
Bagchi, J., Durret, F., Lima Neto, G. B., Paul, S. 2006, Science,
314, 791
\bibitem{Birkinshaw 1999}
Birkinshaw, M. 1999, Phys. Rep., 310, 97
\bibitem{Challinor 1998}
Challinor, A., Lasenby, A. 1998, ApJ, 499, 1
\bibitem{Colafrancesco 2009}
Colafrancesco, S., Prokhorov, D., Dogiel, V. 2009, A\&A, 494, 1
\bibitem{Colafrancesco 2010}
Colafrancesco, S., Marchegiani, P. 2010, A\&A, in press,
[arXiv:astro-ph/0912.2224]
\bibitem{Dubois 2010}
Dubois, Y., Devriendt, J., Slyz, A., Teyssier, R. 2010, MNRAS, in
press, [arXiv:astro-ph/1004.1851]
\bibitem{Giovanelli 1998}
Giovanelli, R., Hayness, M. P., Salzer, J. J., et al. 1998, AJ, 116,
2632
\bibitem{Hansen 2002}
Hansen, S. H., Pastor, S., Semikoz, D. V. 2002, ApJ, 573, L69
\bibitem{Kay 2008}
Kay, S. T., Powell, L. C., Liddle A. R., Thomas, P. A. 2008, MNRAS,
386, 2110
\bibitem{Pointecouteau 1998}
Pointecouteau, E., Giard, M., Barret, D. 1998, A\&A, 336, 44
\bibitem{Prokhorov 2009}
Prokhorov, D. A., Durret, F., Dogiel, V. A., Colafrancesco, S. 2009,
A\&A, 496, 25
\bibitem{Prokhorov 2010}
Prokhorov, D., Antonuccio-Delogu, V., Silk, J. 2010, A\&A, in press,
[arXiv:astro-ph/1006.2564]
\bibitem{Rephaeli 1991}
Rephaeli, Y., Lahav, O. 1991, ApJ, 372, 21
\bibitem{Rephaeli 1995}
Rephaeli, Y. 1995, ApJ, 445, 33
\bibitem{Sarazin 1986}
Sarazin, C. L. 1986, Rev. Mod. Phys., 58, 1
\bibitem{Spergel 2003}
Spergel, D. N., et al. ,2003, ApJS, 148, 175
\bibitem{Sunyaev 1980}
Sunyaev, R. A., Zel'dovich, Ya. B. 1980, ARA\&A, 18, 537
\bibitem{Teyssier 1998}
Teyssier, R. 2002, A\&A, 385, 337
\bibitem{Wright 1979}
Wright, E. L. 1979, ApJ, 232, 348
\end{thebibliography}
\end{document}